\documentclass{ifacconf}

\usepackage{graphicx}      
\usepackage{natbib}        
\usepackage{amsmath}
\usepackage{amssymb}
\begin{document}
\begin{frontmatter}

\title{Sampled-Data Control using Hermite- Obreschkoff Methods with an IDA-PBC Example\thanksref{footnoteinfo}} 

\thanks[footnoteinfo]{Funded by the Deutsche Forschungsgemeinschaft (DFG, German Research Foundation) – 543741503.}

\author[First]{Le Zhang}, 
\author[First]{Paul Kotyczka} 

\address[First]{Technical University of Munich, TUM School of Engineering and Design, Chair of Automatic Control, Garching, Germany (e-mail: le.zhang@tum.de, kotyczka@tum.de).}

\begin{abstract}                
The motivation for this paper is the implementation of nonlinear state feedback control, designed based on the continuous-time plant model, in a sampled control loop under relatively slow sampling. In previous work we have shown that using one-step predictions of the target dynamics with higher order integration schemes, together with possibly higher order input shaping, is a simple and effective way to increase the feasible sampling times until performance degradation and instability occur. In this contribution we present a unifying derivation for arbitrary orders of the previously used Lobatto IIIA collocation and Hermite interpolation schemes through the Hermite-Obreschkoff formula. We derive, moreover, an IDA-PBC controller for a magnetic levitation system, which requires a non-constant target interconnection matrix, and show experimental results.
\end{abstract}


\end{frontmatter}

\section{Introduction}
\label{introduction}
Nonlinear, including passivity-based controls are typically derived in continuous time. In practice, the derived control laws are implemented on digital controllers (computers, programmable logic or micro-controllers), which operate in discrete time with sampled data. Frequently, the continuous control law is implemented directly on the digital controller, in a piecewise constant manner (zero order hold), which corresponds to using the discrete-time model issued from a first order explicit Euler discretization. With sufficiently high sampling frequencies, this simple implementation provides very satisfactory performance. However, for systems with fast dynamics, high sampling frequencies require high-end sensors that can be not only expensive, but also energy consuming. For some cases, it might even be impossible to sample and process the required signals at a sufficiently high rate. In these situations, the controllers suffer from significant performance degradation due to increasing model mismatch, which can lead to destabilization.

To reduce the model mismatch, while also ensuring that the control law designed in continuous time can be directly reused, we initially proposed a second-order discrete-time implementation using the implicit midpoint rule \citep{kotyczka_thoma:2021}. We later extended the approach to $s$-stage Gauss-Legendre collocation with higher order input shaping \citep{kotyczka_martens_lefevre:2021}. In \cite{kotyczka_2023}, we discussed the sampled control implementation with cubic Hermite interpolation or 3-stage Lobatto IIIA collocation, which are equivalent symmetric methods, and could potentially reduce the discontinuity of the control signal at the sampling instants. In contrast to the higher-order-hold elements in our work, efforts have been made to derive piecewise constant control signal based on series expansion of continuous-time IDA-PBC \citep{tiefensee:2010}.

The first topic of this contribution is to generalize the two methods implemented in \cite{kotyczka_2023} towards Hermite-Obreschkoff (HO) methods. We show that Lobatto IIIA collocation of \emph{arbitrary} order can be derived using the HO formula, and that the derivation gives rise to a specific set of interpolation splines of the corresponding order. We argue that this interpolation scheme is equivalent to Hermite interpolation after some mathematical manipulation. We further present the use of this interpolation scheme in practical implementation on a magnetic levitation experiment, where the passivity-based control design with the IDA-PBC method \citep{ortega2002} requires a non-constant interconnection matrix due to the form of the distance-dependent inductance function.

In Section \ref{ho_methods}, we first introduce the HO formula under the setting of solving an initial value problem (IVP). We then show respectively the derivation of Lobatto IIIA collocation and the corresponding interpolation schemes using this formula. In Section \ref{implementation}, we recall the higher-order input shaping introduced in previous works. As a workaround to prevent strongly oscillating inputs due to their polynomial shape and modeling inaccuracies, we propose how to convert the control to a piecewise constant (zero order hold) signal. In Section \ref{experiment}, we present the novel IDA-PBC controller for the magnetic levitation system and discuss some experimental observations regarding its discrete-time implementations.

\section{Solving Initial Value Problems using Hermite-Obreschkoff Methods}
\label{ho_methods}
We consider a time-varying initial value problem of the form
\begin{equation}
	\label{eq:ivp}
	\dot{x} (t) = f (t, x(t)), \quad x(t_{k}) = x_{k},
\end{equation}
on an arbitrary time interval $I_{k} = [t_{k}, t_{k+1}]$, $t_{k} = kh$, $h = const.$, $k \in \mathbb{N}_{0}$ (equidistant sampling). The goal is to numerically approximate its solution $x(t_{k+1})$ for a given $x(t_{k})$. Such a one-step prediction will be used later for the sampled control implementation.

\subsection{Hermite-Obreschkoff Formula}
\label{ho_formula}
\cite{obreschkoff:1940} suggested the following new quadrature formula based on Taylor series and Cesàro summation \citep{cesaro1890}:
\begin{equation}
	\label{eq:ho_formula}
	\sum_{j=0}^{n} A_{n-j}^{m} \frac{(\tau h)^{j}}{j!} \frac{\mathrm{d}^{j}}{\mathrm{d} t^{j}} x(t_{k}) = \sum_{j=0}^{m} A_{n}^{m-j} \frac{(-\tau h)^{j}}{j!} \frac{\mathrm{d}^{j}}{\mathrm{d} t^{j}} \tilde{x}(t_{k}+\tau h),
\end{equation}
where $x(t)$ is assumed to be $(n+m+1)$-times differentiable, $n,m \in \mathbb{N}_{0}$, $\tau \in [0,1]$, and $A_{n}^{m} = \binom{n+m}{n} = \frac{(n+m)!}{n!m!}$. The ``$\sim$'' sign on the right-hand-side of (\ref{eq:ho_formula}) indicates that $\tilde x(t_k + \tau h)$ is an approximation, which has a local error of $\mathcal{O}(h^{n+m+1})$.

We choose the Hermite-Obreschkoff formula as a versatile starting point, as it is a very general formulation of the approximation of functions, and has been widely advocated for the computation of integrals as well as the solution of differential equations \citep{hairer:1993}.

The general idea of HO methods is to combine the HO formulas defined by different choices of the integer pair $(n,m)$, while the local errors remain the same order, i.e., $n+m$ stays constant. This consistency allows the local error of the resulting method to have the same order, and avoids confusion in the truncation error. We call the resulting numerical integration scheme from such a combination an HO method.

\subsection{Lobatto IIIA Collocation}
\label{LIIIA}
We now show that the Lobatto IIIA collocation schemes are members of the HO methods, in the sense that they can be derived through the combination of HO formulas.

The collocation points used in Lobatto IIIA methods are the zeros of the shifted Legendre polynomial
\begin{equation}
	\label{eq:legendre}
	\frac{\mathrm{d}^{s-2}}{\mathrm{d} x^{s-2}} (x^{s-1}(x-1)^{s-1}),
\end{equation}
where the integer $s \ge 2$ is the number of collocation points (the number is not smaller than 2 since there are at least 0 and 1), see \cite{hairer:2006}. We denote these points as $c_{i}$, $i=1,\ldots,s$ (where $c_{1} = 0$, $c_{s} = 1$). By substituting $\tau$ in (\ref{eq:ho_formula}) with each $c_{i}$, the Lobatto IIIA collocation schemes can be derived.

\begin{thm}
	If the nodes $c_{i}$, $i=1,\ldots,s$, in which the integer $s \ge 2$, are the zeros of (\ref{eq:legendre}), then the Hermite-Obreschkoff method using the pairs $(s,0)$ and $(s-1,1)$ is equivalent to the $s$-stage Lobatto IIIA method.
\end{thm}

\begin{pf}
	For convenience, we first introduce some notation. For the state variable, we denote
	\begin{equation*}
		x_{k} := x (t_{k}), \quad \tilde{x}_{k,i} := \tilde{x} (t_{k} + c_{i}h).
	\end{equation*}
	For the time derivatives, we use the vector field $f$:
	\begin{equation*}
		\begin{array}{ll}
			&\overset{(j-1)}{\tilde{f}_{k,i}} := \frac{\mathrm{d}^{j-1}}{\mathrm{d} t^{j-1}} f(t_{k} + c_{i}h, \tilde{x} (t_{k} + c_{i}h)) = \frac{\mathrm{d}^{j}}{\mathrm{d} t^{j}} \tilde{x} (t_{k} + c_{i}h),\\
			&\overset{(j-1)}{f_{k}} := \frac{\mathrm{d}^{j-1}}{\mathrm{d} t^{j-1}} f(t_{k}, x(t_{k})) = \frac{\mathrm{d}^{j}}{\mathrm{d} t^{j}} x(t_{k}),\quad i,j = 1, \ldots, s.
		\end{array}		
	\end{equation*}
	For now we do not wish to involve time derivatives of $f$ at the nodes, i.e., $m$ cannot be greater than $1$. Choosing the integer pair $(n,m)$ to be $(s,0)$ and $(s-1,1)$ respectively, (\ref{eq:ho_formula}) yields two ``master'' equations
	\begin{subequations}
		\label{eq:master_equations}
		\begin{align}
			\label{eq:(s,0)}
			\tilde{x}_{k,i} &= x_{k} + \sum_{j=1}^{s} \frac{(c_{i}h)^{j}}{j!} \, \overset{(j-1)}{f_{k}}, \\
			\label{eq:(s-1,1)}
			s\tilde{x}_{k,i} - c_{i}h\tilde{f}_{k,i} &= sx_{k} + \sum_{j=1}^{s-1} (s-j) \frac{(c_{i}h)^{j}}{j!} \, \overset{(j-1)}{f_{k}}. 
		\end{align}
	\end{subequations}
	 With the nodes $c_{i}, i = 1,\ldots,s$, (\ref{eq:master_equations}) gives $2s$ equations in total. But we notice that for $c_{1} = 0$, both (\ref{eq:(s,0)}) and (\ref{eq:(s-1,1)}) give the same equation $\tilde{x}_{k,1} = x_{k}$, which indicates $\tilde{f}_{k,1} = f_{k}$. Therefore, we have in fact $2s-1$ equations, which can be written in the compact form
	\begin{equation}
		\label{eq:compact_equations}
		\begin{split}
			&\begin{bmatrix}
				I_{s\times s} &P_{s\times (s-1)} \\
				O_{(s-1)\times s} &Q_{(s-1) \times (s-1)}
			\end{bmatrix}
			\begin{bmatrix}
				\tilde{X} \\
				\Delta_{k}
			\end{bmatrix}  \\
			&\qquad \qquad \qquad \qquad =\begin{bmatrix}
				I_{s\times s} &M_{s \times s} \\
				O_{(s-1)\times s} &N_{(s-1) \times s}
			\end{bmatrix}
			\begin{bmatrix}
				X_{k} \\
				h\tilde{F}
			\end{bmatrix},
		\end{split}
	\end{equation}
	where $\tilde{X}, \Delta_{k}, X_{k}, \tilde{F}$ are matrices containing the variables 
	\begin{equation}
		\label{eq:variables}
		\begin{array}{lll}
			\tilde{X} &:= [\tilde{x}_{k,1}, \ldots, \tilde{x}_{k,s}]^{T}, &\Delta_{k} := [\dot{f_{k}}, \ldots, \overset{(s-1)}{f_{k}}]^{T}, \\
			X_k &:= [x_{k}, \ldots, x_{k}]^{T}, &\tilde{F} := [\tilde{f}_{k,1}, \tilde{f}_{k,2}, \ldots, \tilde{f}_{k,s}]^{T},
		\end{array}
	\end{equation}
	and $P,Q,M,N$ are constant matrices, whose dimensions are indicated by their subscripts.
	
	According to (\ref{eq:master_equations}), the elements of the matrices $P,Q$ are verified to be
	\begin{align*}
		P_{i,j} &= -\frac{(c_{i}h)^{j+1}}{(j+1)!}, \, i=1,\ldots,s, \, j = 1,\ldots,s-1, \\
		Q_{i,j} &= -\frac{1}{s} \frac{(c_{i+1}h)^{j+1}}{j!}, \, i,j = 1,\ldots,s-1.		
	\end{align*}
	The matrices $M,N$ turn out to be
	\begin{equation*}
		M = \begin{bmatrix}
			c_{1} &0 &\cdots &0 \\
			\vdots &\vdots & &\vdots \\
			c_{s} &0 &\cdots &0
		\end{bmatrix}, \,
		N = \frac{1}{s}\begin{bmatrix}
			c_{2} &-c_{2} &0 &\cdots &0 \\
			c_{3} &0 &-c_{3} &\cdots &0 \\
			\vdots &\vdots &\vdots &\ddots &\vdots \\
			c_{s} &0 &0 &\cdots &-c_{s}
		\end{bmatrix}.
	\end{equation*}
	Now that (\ref{eq:compact_equations}) is well defined, we observe that if the nodes $c_{2},\ldots,c_{s}$ are distinct from each other (which is true by their definition), and the sampling time $h$ is non-zero, then $Q$ is invertible, and (\ref{eq:compact_equations}) yields the unique solution
	\begin{equation}
		\label{eq:solution}
		\tilde{X} = X_{k} + h A_{s} \tilde{F},
	\end{equation}
	in which the coefficient matrix $A_{s}$ is defined as
	\begin{equation}
		\label{eq:A}
		A_{s}:=M-PQ^{-1}N.
	\end{equation}
	For a given integer $s$, (\ref{eq:solution}) gives the exactly same formulation as the $s$-stage Lobatto IIIA collocation shown in \cite{hairer:2006}.~~~~~~~~~~~~~~~~~~~~~~~~~~~~~~~~~~~~~~~~~~~~~~~\qed
	\end{pf}

\begin{rem}
	We have defined the matrices as in (\ref{eq:variables}) to avoid the Kronecker product. In terms of the stacked column vectors of $\text{vec}(\tilde{X}^{T})$, (\ref{eq:solution}) would read
	\begin{equation}
		\text{vec}(\tilde{X}^{T}) = \text{vec}(\tilde{X}_{k}^{T}) + h (A_{s} \otimes I) \text{vec}(\tilde{F}^{T}).
	\end{equation}
\end{rem}

\begin{rem}
	By setting the pair $(n,m)$ to be $(s,0)$ and $(s-1,1)$, we ensure that the resulting HO method is an approximation of order $s$. However, the corresponding $s$-stage Lobatto IIIA collocation is an approximation of order $2s-2$ \citep{hairer:2006}. Therefore, the HO method only provides a very conservative evaluation on the approximation order. For an accurate evaluation of the approximation order, we refer to the ``rooted-tree-type'' theory presented in \cite{jay1994}, and the proof of \emph{superconvergence} presented in \cite{hairer:2006}.
\end{rem}

\subsubsection{Example.} For $s=2$ and $s=3$, the nodes are $\left\lbrace 0,1\right\rbrace$ and $\left\lbrace0, \frac{1}{2}, 1\right\rbrace$ respectively. Using (\ref{eq:A}), the coefficient matrices are
\begin{equation}
	\label{eq:A_2_A_3}
	A_{2} = \begin{bmatrix}
		\vspace{0.5ex}
		0 &0 \\
		\vspace{0.5ex}
		\frac{1}{2} &\frac{1}{2}
	\end{bmatrix},\quad
	A_{3} = \begin{bmatrix}
		\vspace{0.5ex}
		0 &0 &0 \\
		\vspace{0.5ex}
		\frac{5}{24} &\frac{1}{3} &-\frac{1}{24} \\
		\vspace{0.5ex}
		\frac{1}{6} &\frac{2}{3} &\frac{1}{6}
	\end{bmatrix},
\end{equation}
which are indeed identical to those of the 2- and 3-stage Lobatto IIIA collocation written as Runge–Kutta schemes.

\subsection{Hermite Interpolation}
\label{interpolation}
The derivation of Lobatto IIIA collocation only makes use of HO formulas with the pairs $(s,0)$ and $(s-1,1)$. Now we investigate other pairs with the same order.

Firstly, we assume the pair is now $(s-m,m)$, $m=2$. This pair exists since $s\ge2$. A new ``master'' equation for this pair is
\begin{equation}
	\label{eq:(s-m,m)}
	\begin{split}
		&A_{s-m}^{m}s\tilde{x}_{k,i} - A_{s-m}^{m-1}c_{i}h\tilde{f}_{k,i} + \sum_{j = 2}^{m} A_{s-m}^{m-j} \frac{(-c_{i}h)^{j}}{j!} \overset{(j-1)}{\tilde{f}_{k,i}} \\
		&= A_{s-m}^{m} x_{k} + \sum_{j=1}^{s-m} A_{s-m-j}^{m} \frac{(c_{i}h)^{j}}{j!} \, \overset{(j-1)}{f_{k}},
	\end{split}
\end{equation}
where we pull out the first two terms of the sum on the left, and the first term on the right for clarity. (\ref{eq:(s-m,m)}) gives $s-1$ new equations with $c_{i}$, $i = 2,\ldots,s$ (with $c_{1} = 0$ we get $\tilde{x}_{k,1}=x_k$ again). Comparing (\ref{eq:(s-m,m)}) with (\ref{eq:(s-1,1)}), the main difference is the new unknown variables
\begin{equation*}
	\overset{(m-1)}{\tilde{f}_{k,2}},\, \ldots, \, \overset{(m-1)}{\tilde{f}_{k,s}}.
\end{equation*}
Since all the other variables are determined by (\ref{eq:compact_equations}), we can express these new unknowns as functions of $\tilde{F}$, and move them all to the right-hand side. (\ref{eq:(s-m,m)}) thus becomes
\begin{equation}
	\label{eq:(s-m,m):solution}
	\frac{(-h)^{m}}{m!}\begin{bmatrix}
		c_{2}^{m} & & \\
		 &\ddots & \\
		 & &c_{s}^{m}
	\end{bmatrix}
	\begin{bmatrix}
		(\overset{(m-1)}{\tilde{f}_{k,2}})^{T} \\
		\vdots \\
		(\overset{(m-1)}{\tilde{f}_{k,s}})^{T}
	\end{bmatrix}
	 = E_{(s-1)\times s} \tilde{F},
\end{equation}
where $E$ represent some constant matrix dependent on $h$ and $c_{i}$. Since $h$ is non-zero, and the nodes are distinct from each other, these $s-1$ new unknowns can also be uniquely solved for. Should $s$ be greater than 2, the exactly same process can be carried out for each $m = 3, \ldots,s$.

As the second step, we define the following matrices
\begin{equation*}
	\dot{\tilde{F}} = [\dot{\tilde{f}}_{k,1}, \cdots, \dot{\tilde{f}}_{k,s}]^{T}, \quad \cdots,\quad
	\overset{(s-1)}{\tilde{F}} = [\overset{(s-1)}{\tilde{f}_{k,1}}, \cdots , \overset{(s-1)}{\tilde{f}_{k,s}}]^{T},
\end{equation*}
in which $\overset{(j)}{\tilde{f}_{k,1}} = \overset{(j)}{f_{k}},\, j = 1,\ldots,s-1,$
is determined by (\ref{eq:compact_equations}), and the rest can be solved for with (\ref{eq:(s-m,m):solution}). Together with $\tilde{X}$, we have $s^{2}$ variables to be solved for.

On the other hand, using all the pairs $(n,m)$ satisfying $n+m=s$, and the $s$ nodes, the HO formula yields $s^{2}$ equations in total (removing the identical equations). Through the process above (including Section \ref{LIIIA}), we show that these $s^{2}$ equations are actually \emph{linearly independent} with respect to the $s^{2}$ variables. Therefore, they can be solved uniquely and also efficiently.

We can organize all the solutions in a uniform manner:
\begin{equation}
	\label{eq:solution_all}
	\begin{array}{ccrl}
		\overset{(s-1)}{\tilde{F}} &=& \frac{1}{h^{s-1}} &D_{s}^{(s-1)} \tilde{F}, \\
		&\vdots& & \\
		\dot{\tilde{F}} &=& \frac{1}{h} &D_{s}^{(1)} \tilde{F}, \\
		\tilde{F} &=& &D_{s}^{(0)} \tilde{F}, \\
		\tilde{X} &=& X_{k} + h&A_{s} \tilde{F},
	\end{array}
\end{equation}
where $A_{s}$ is defined by (\ref{eq:A}), $D_{s}^{(i)}$, $i=0,\ldots,s-1$ are some constant coefficient matrices, among which $D_{s}^{(0)}$ is always the identity matrix, and the rest are computed through (\ref{eq:compact_equations}) and (\ref{eq:(s-m,m):solution}). Note that the second last equation is trivial, but deliberately added for the sake of uniformity, as $\tilde{F}$ is not unknown. From another point of view, this equation actually represents the collocation condition.

Two observations can be made after (\ref{eq:solution_all}) is computed:
\begin{itemize}
	\item All rows of $D_{s}^{(s-1)}$ are identical, suggesting that $\frac{\mathrm{d}^{s-1}}{\mathrm{d}t^{s-1}}f$ is approximated by a constant value across the time interval $I_{k}$.
	\item Each equation is the time integral of the equation above it.
\end{itemize}

For the final step, we introduce the following notation to represent the approximations across the time interval $I_{k}$
\begin{equation*}
	\begin{array}{rl}
		\tilde{x}_{k+\tau} &:= \tilde{x}(t_{k}+\tau h), \\
		\overset{(j-1)}{\tilde{f}_{k+\tau}} &:= \frac{\mathrm{d}^{j-1}}{\mathrm{d} t^{j-1}} f(t_{k} + \tau h, \tilde{x} (t_{k} + \tau h)), \quad j=1,\ldots,s,
	\end{array}
\end{equation*}
where $\tau \in [0,1]$ is the normalized time. Note that $\mathrm{d}t = h\mathrm{d}\tau$, and therefore $\frac{\mathrm{d}}{\mathrm{d} t} = \frac{1}{h} \frac{\mathrm{d}}{\mathrm{d} \tau}$.

Now we express all the approximations involved in (\ref{eq:solution_all}) using $\tau$:
\begin{equation}
	\label{eq:interpolation_all}
	\begin{array}{ccrl}
		\overset{(s-1)}{\tilde{f}_{k+\tau}} &=& \frac{1}{h^{s-1}} \tilde{F}^{T} & \frac{\mathrm{d}^{s}}{\mathrm{d} \tau^{s}} H_{s} (\tau) , \\
		&\vdots& & \\
		\vspace{0.5ex}
		\dot{\tilde{f}}_{k+\tau} &=& \frac{1}{h} \tilde{F}^{T} &\frac{\mathrm{d}^{2}}{\mathrm{d} \tau^{2}} H_{s} (\tau), \\
		\vspace{0.5ex}
		\tilde{f}_{k+\tau} &=& \tilde{F}^{T} &\frac{\mathrm{d}}{\mathrm{d} \tau} H_{s} (\tau), \\
		\vspace{0.5ex}
		\tilde{x}_{k+\tau} &=& x_{k} + h\tilde{F}^{T} & H_{s} (\tau),
	\end{array}
\end{equation}
where $H_{s} (\tau)$ is a vector function of dimension $s$, of which each element is a polynomial of $\tau$.

Based on the observations above, we can state that $(\mathrm{d}^{s}/\mathrm{d} \tau^{s}) H_{s} (\tau)$ is a constant vector, and each element of $H_{s} (\tau)$ is a polynomial of order $s$. Thus $H_{s} (\tau)$ contains $s(s+1)$ coefficients to be defined. Matching (\ref{eq:interpolation_all}) at the nodes with (\ref{eq:solution_all}), yields
\begin{equation*}
	\begin{array}{cl}
		\vspace{0.4ex}
		\begin{bmatrix}
			\frac{\mathrm{d}^{i}}{\mathrm{d} \tau^{i}} H_{s} (c_{1}) &\cdots & \frac{\mathrm{d}^{i}}{\mathrm{d} \tau^{i}} H_{s} (c_{s})
		\end{bmatrix}
		&= (D_{s}^{(i-1)})^{T}, \, i = 1,\ldots,s,\\
		\begin{bmatrix}
			\,\,\,\,\, H_{s} (c_{1}) \quad &\cdots & \quad H_{s} (c_{s}) \,\,\,\,\,
		\end{bmatrix} &= A_{s}^{T}.
	\end{array}
\end{equation*}
which defines these coefficients and gives an interpolation scheme, which we formulate below.

\begin{defn}
	If the nodes $c_{i}$, $i=1,\ldots,s$, in which the integer $s \ge 2$, are the zeros of (\ref{eq:legendre}), then the Hermite-Obreschkoff method using the pairs $(n,m)$ satisfying
	\begin{equation*}
		n+m = s, \quad n,m \in \mathbb{N}_{0},
	\end{equation*}
	is an $s$-order Hermite interpolation scheme of the form
	\begin{equation}
		\label{eq:interplation}
		\tilde{x}(t_{k} + \tau h) = x_{k} + h\tilde{F}^{T} H_{s} (\tau),\quad \tau \in [0,1],
	\end{equation}
	where $H_{s}(\tau)$ is a set of interpolation splines of order $s$.
\end{defn}

\subsubsection{Example.}
For $s = 3$, it is computed that
\begin{equation*}
	D_{3}^{(2)} = \begin{bmatrix}
		4 &-8 &4 \\
		4 &-8 &4 \\
		4 &-8 &4 \\
	\end{bmatrix},\quad
	D_{3}^{(1)} = \begin{bmatrix}
		-3 &4 &-1 \\
		-1 &0 &1 \\
		1 &-4 &3 \\
	\end{bmatrix},
\end{equation*}
and $D_{3}^{(0)} = I_{3}$, $A_{3}$ is defined in (\ref{eq:A_2_A_3}). The splines turn out to be
\begin{equation*}
	H_{3}(\tau) = \begin{bmatrix}
		\frac{2}{3} \tau^{3} - \frac{3}{2} \tau^{2} + \tau \\
		-\frac{4}{3} \tau^{3} + 2\tau^{2} \\
		\frac{2}{3} \tau^{3} - \frac{1}{2} \tau^{2}
	\end{bmatrix},\quad \tau \in [0,1].
\end{equation*}
From the last row of (\ref{eq:solution}), we have
\begin{equation}
	\label{eq:f_k_2}
	\tilde{f}_{k,2} = -\frac{3}{2h} x_{k} + \frac{3}{2h} \tilde{x}_{k,3} -\frac{1}{4} \tilde{f}_{k,1} - \frac{1}{4} \tilde{f}_{k,3}.
\end{equation}
Substituting (\ref{eq:f_k_2}) into (\ref{eq:interplation}), and replacing $\tilde{x}_{k,3},\tilde{f}_{k,3}$ with $\tilde{x}_{k+1},\tilde{f}_{k+1}$ (since $c_{3}=1$), (\ref{eq:interplation}) becomes
\begin{equation*}
	\begin{array}{rl}
		\vspace{0.5ex}
		\tilde{x}(t_{k} + \tau h) =& \begin{bmatrix}
			x_{k} &\tilde{x}_{k+1}
		\end{bmatrix} \begin{bmatrix}
		2\tau^{3} - 3\tau^{2} + 1 \\
		-2\tau^{3} + 3\tau^{2}
		\end{bmatrix} \\
		\vspace{0.5ex}
		&+h\begin{bmatrix}
			f_{k} &\tilde{f}_{k+1}
		\end{bmatrix} \begin{bmatrix}
		\tau^{3}-2\tau^{2}+\tau \\
		\tau^{3} - \tau^{2}
		\end{bmatrix},
	\end{array}
\end{equation*}
which is the cubic Hermite interpolation. Similarly, the quintic Hermite interpolation can be derived.

In \cite{kotyczka_2023}, it was proven that numerical integration based on the cubic Hermite interpolation is equivalent to the 3-stage Lobatto IIIA collocation. In this contribution, we set out from a different starting point, and find out that both of these methods, for arbitrary number of stages $s$, are members of the HO methods, and they both actually stem from solving the same set of HO formulas defined by the same nodes (zeros of (\ref{eq:legendre})).

\begin{prop}
	Numerical integration (solution of IVP (\ref{eq:ivp})) based on the $s$-stage Lobatto IIIA collocation is equivalent to the Hermite interpolation of order $s$.
\end{prop}

\section{Higher-Order Discrete-Time Control Implementation}
\label{implementation}
For the implementation in a sampled control loop of the HO methods introduced above, we first recall briefly the input shaping method as described and presented in simulations in \cite{kotyczka_2023}, and then propose a technique to generate piecewise constant input signals by utilizing the Hermite interpolation schemes.

\subsection{Input Shaping Based on Lagrange Interpolation}
For a given control system
\begin{equation}
	\label{eq:system}
	\dot{x}(t) = f(t, x(t), u(t)), \quad x(0) = x_{0},
\end{equation}
and a state feedback law designed in continuous time $u(t) = r(t,x(t))$, the desired (subscript ``d'') closed-loop system is
\begin{equation}
	\label{eq:system_cl}
		\vspace{0.5ex}
		\dot{x} (t) = f(t, x(t) ,r(t,x(t))) = f_{d}(t,x(t)).
\end{equation}
We denote the approximated solution of (\ref{eq:system_cl}) at the nodes of the interval $I_k$ as $(\tilde{x}_{d})_{k,i}$, $i = 1,\ldots,s$. The control input signal on the sampling interval $I_{k}$ is shaped using an $(s-1)$-order-hold element:
\begin{subequations}
	\label{eq:shaped_input}
	\begin{gather}
		u_{k,i} = r(t_k+c_{i}h, (\tilde{x}_{d})_{k,i}),\quad i = 1,\ldots,s,\\
		u(t_k + \tau h) = \sum_{i=1}^{s} \ell_{i}^{(s-1)}(\tau) u_{k,i},
	\end{gather}
\end{subequations}
where $\ell_{i}^{(s-1)}(\tau)$ are Lagrange interpolation polynomials of degree $s-1$, defined as $\ell_{i}^{(s-1)}(\tau) = \prod_{j = 1,j \ne i}^{s} \frac{\tau-c_{j}}{c_{i}-c_{j}}$, $i = 1,\ldots,s$.

It was proven that, for sufficiently small $h$, if $(\tilde{x}_{d})_{k,i}$ are the stage values of an order $p$ accurate numerical solution of (\ref{eq:system_cl}) on the sampling interval $I_k$ (a one-step prediction of the target dynamics), then the trajectories of (\ref{eq:system}) with the shaped input (\ref{eq:shaped_input}) are approximations of order $p$ of the desired closed-loop dynamics, see \cite{kotyczka_2023}.

\subsection{Conversion to Piecewise Constant Input}
The input shaping method using an $(s-1)$-order hold element is theoretically sound, and provides very good performance. However, it requires the input signal to follow a high-order polynomial within one sampling interval, which poses more challenges to the actuators. In practice and in other control implementations (such as model predictive control), it is often the case that the actuators have the same operating rate as the sampling rate, and piecewise constant control input are desired.

To compute an appropriate value of constant input $u_k$ on $I_k$, such that the states can reach the same value as desired at the end of the interval, we propose to compute $u_k$ as the solution of the following optimization problem:
\begin{subequations}
	\label{eq:optimization}
	\begin{align}
		&\quad\min_{u_k} \left\| \sum_{i=1}^{s} b_{i}\Big(f(t_{k,i}, \tilde{x}_{k,i}, u_{k}) -f_{d}(t_{k,i},(\tilde{x}_{d})_{k,i}) \Big) \right\|  \\ \nonumber
		&\text{such that for}\,\, i=1,\ldots,s:\\ 
		&\qquad\qquad \tilde{x}_{k,i} = x_k + h\sum_{j=1}^{s}a_{ij}f(t_{k,i}, \tilde{x}_{k,i}, u_{k}) \\
		&\qquad\quad(\tilde{x}_{d})_{k,i} = x_k + h\sum_{j=1}^{s}a_{ij} f_{d}(t_{k,i},(\tilde{x}_{d})_{k,i})
	\end{align}
\end{subequations}
where $a_{ij}$ are the elements of $A_{s}$ and $b_{i}$ the elements of $B:=(H_{s}(1))^{T}$.

From (\ref{eq:interplation}), it can be shown that the optimization problem (\ref{eq:optimization}) minimizes the difference $\left\|\tilde{x}_{k+1} - (\tilde{x}_{d})_{k+1} \right\|$. This conversion addresses the problem of practical implementation, but increases the approximation error of the embedded HO method, which will be shown in the following experiments.

\section{Sampled IDA-PBC for the Magnetic Levitation System}
\label{experiment}
Finally, as validation, we implement Lobatto IIIA collocation and Hermite interpolation on the magnetic levitation system (MagLev) test bench, using both the higher-order shaped input and the piecewise constant input. First, we show the continuous-time control design of the MagLev using IDA-PBC, then we present the experimental results of its discrete-time implementation using the above mentioned HO methods.

\begin{figure}
	\begin{center}
		\includegraphics[width=6.5cm]{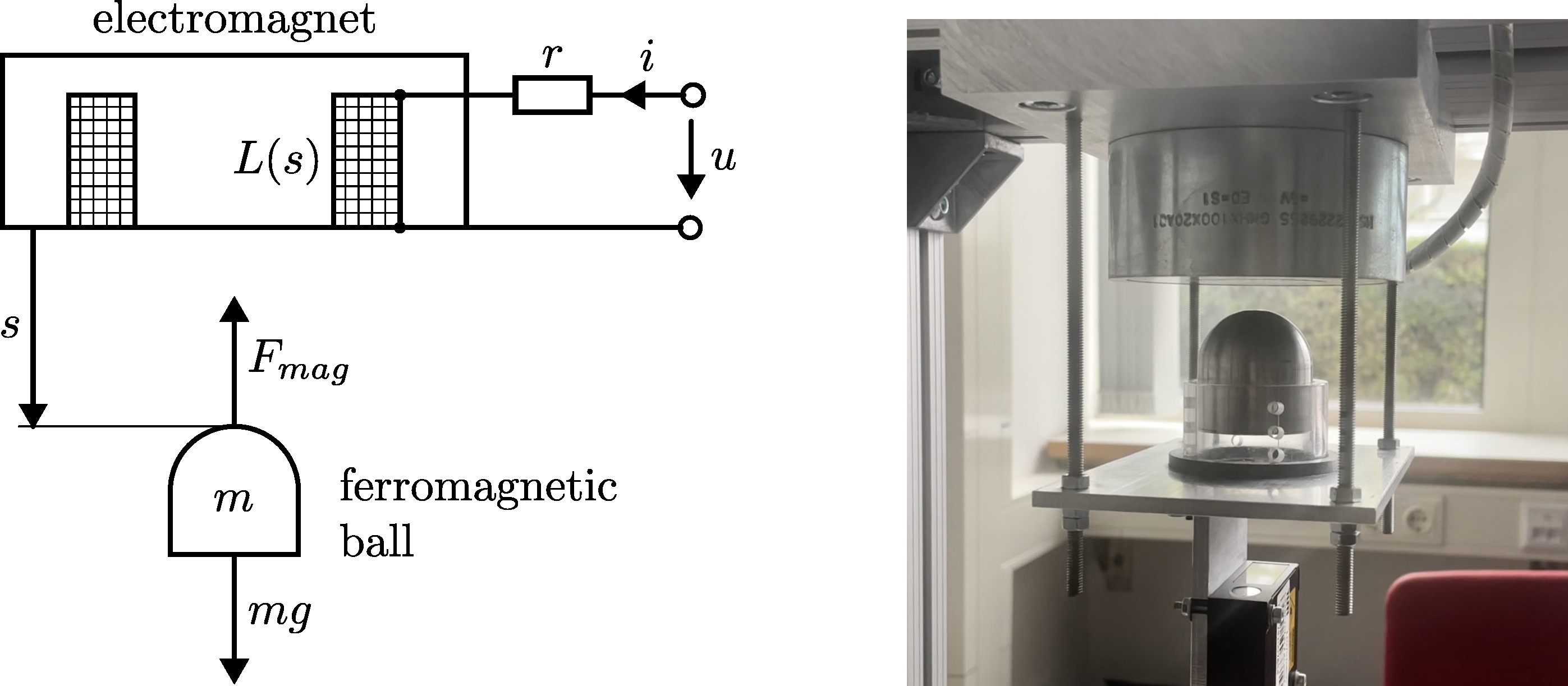}
		\caption{Sketch and photo of the MagLev test bench.}
		\label{fig:system}
	\end{center}
\end{figure}

\subsection{IDA-PBC Design}
The MagLev system is depicted in Fig.~\ref{fig:system}. Its continuous time dynamics can be described as follows:
\begin{equation}
	\label{eq_maglev}
	\begin{array}{rl}
		\vspace{0.5ex}
		\dot{s} &= \frac{p}{m}, \\
		\vspace{0.5ex}
		\dot{p} &= \frac{1}{2}L^{\prime}(s) i^{2} + mg, \\
		\vspace{0.5ex}
		\dot{i} &= -\frac{1}{L(s)}(r + L^{\prime}(s)\frac{p}{m})i + \frac{1}{L(s)} u.
	\end{array}
\end{equation}
The states $s,p,i$ are the distance between the ball and the magnet, the momentum of the ball, and the electric current (always positive), the control input $u$ is the voltage applied to the coil, and $L(s)$ is the inductance of the electromagnet identified as
\begin{equation*}
	L(s) = L_\infty + \frac{a}{(bs+1)^{3}}, \quad L^{\prime}(s) = -\frac{3ab}{(bs+1)^{4}}.
\end{equation*}
This more precise inductance characteristic differs from the one used in \cite{kotyczka_martens_lefevre:2021}, and requires a state dependent target interconnection matrix for IDA-PBC. The identified system parameters are presented in Table~\ref{tb:parameters}.

The system has a strict-feedback form \citep{khalil2002}, meaning that a fictitious control law $i^2 = \phi (s,p)$ can be found for the subsystem $[s , p]^{T}$. By the state transformation $z := i^{2} - \phi (s,p)$, the system becomes
\begin{equation}
	\label{eq_maglev_shifted}
	\underbrace{\begin{bmatrix}
		\dot{s} \\
		\dot{p} \\
		\dot{z}
	\end{bmatrix}}_{\dot{\hat{x}}} = \underbrace{\begin{bmatrix}
		\vspace{0.5ex}
		\frac{p}{m} \\
		\vspace{0.5ex}
		mg + \frac{L^{\prime}(s)}{2} (z+\phi) \\
		\vspace{0.5ex}
		f(s,p,z)
	\end{bmatrix}}_{\hat{f}(\hat{x})} + \underbrace{\begin{bmatrix}
		0 \\
		0 \\
		g (s,p,z)
	\end{bmatrix}}_{\hat{G}(\hat{x})}u,
\end{equation}
where
\begin{equation*}
	\begin{array}{rl}
		f(s,p,z) &=  -\frac{2}{L(s)}(r + L^{\prime}(s)\frac{p}{m})(z+\phi) - \dot{\phi},\\
		g(s,p,z) &= \frac{2\sqrt{z+\phi}}{L(s)}.
	\end{array}
\end{equation*}
For IDA-PBC, (\ref{eq_maglev_shifted}) with the control law $u = \hat{r} (\hat{x})$ (to be determined), is matched with the target dynamics $\dot{\hat{x}} = (\hat{J}_{d}-\hat{R}_{d}) \nabla \hat{H}_d (\hat{x})$, with the fictitious control law designed as
\begin{equation*}
	\phi(s,p) = \frac{2}{L^{\prime}(s)} (-C(s-s^{\ast}) - k_{1}\frac{p}{m} - mg).
\end{equation*}
It can be verified that the matching conditions are satisfied with the interconnection and damping matrices
\begin{equation*}
	\hat{J}_d = \begin{bmatrix}
		0 &1 &0 \\
		-1 &0 &\frac{L^{\prime}(s)}{2} \\
		0 &-\frac{L^{\prime}(s)}{2} &0
	\end{bmatrix}, \quad
	\hat{R}_d = \begin{bmatrix}
		0 &0 &0 \\
		0 &k_{1} &0 \\
		0 &0 &{k_{2}}
	\end{bmatrix},
\end{equation*}
and the closed-loop energy function
\begin{equation}
	\hat H_{d} (\hat{x}) = \frac{p^{2}}{2m} + \frac{C}{2} (s-s^{\ast})^{2} + \frac{1}{2} z^{2},
\end{equation}
where $C,k_{1},k_{2} > 0$, and $s^{\ast}$ is the desired position of the ball. The final control law is then obtained as
\begin{equation}
	\label{eq:ida_pbc_shifted}
	\hat{r}(s,p,z) = (\hat{G}^{T}\hat{G})^{-1}\hat{G}^{T}((\hat{J}_{d}-\hat{R}_{d})\nabla \hat{H}_d - \hat{f}).
\end{equation}
Transforming (\ref{eq:ida_pbc_shifted}) back to the original coordinates yields
\begin{equation*}
	r(s,p, i) = \hat{r}(s,p,i^{2}-\phi (s,p)).
\end{equation*}

\begin{table}[bp]
	\begin{center}
		\caption{System parameters}\label{tb:parameters}
		\begin{tabular}{cccc}
			Name & Symbol & Value & Unit \\ \hline
			Mass of the ball & $m$ & $85.9\times 10^{-3}$ & $\mathrm{kg}$ \\
			Gravitational acceleration & $g$ & $9.81$ & $\mathrm{m/s^2}$ \\
			Resistance & $r$ & $2.1512$ & $\mathrm{\Omega}$ \\
			Inductance parameters & $L_{\infty}$ & $54.9\times 10^{-3}$ & $\mathrm{H}$ \\
			& $a$ & $15\times 10^{-3}$ & $\mathrm{H}$ \\
			& $b$ & $50.4131$ & $\mathrm{m^{-1}}$ \\ \hline
		\end{tabular}
	\end{center}
\end{table}

\subsection{Experimental Results}
\begin{figure}
	\begin{center}
		\includegraphics[width=6.8cm]{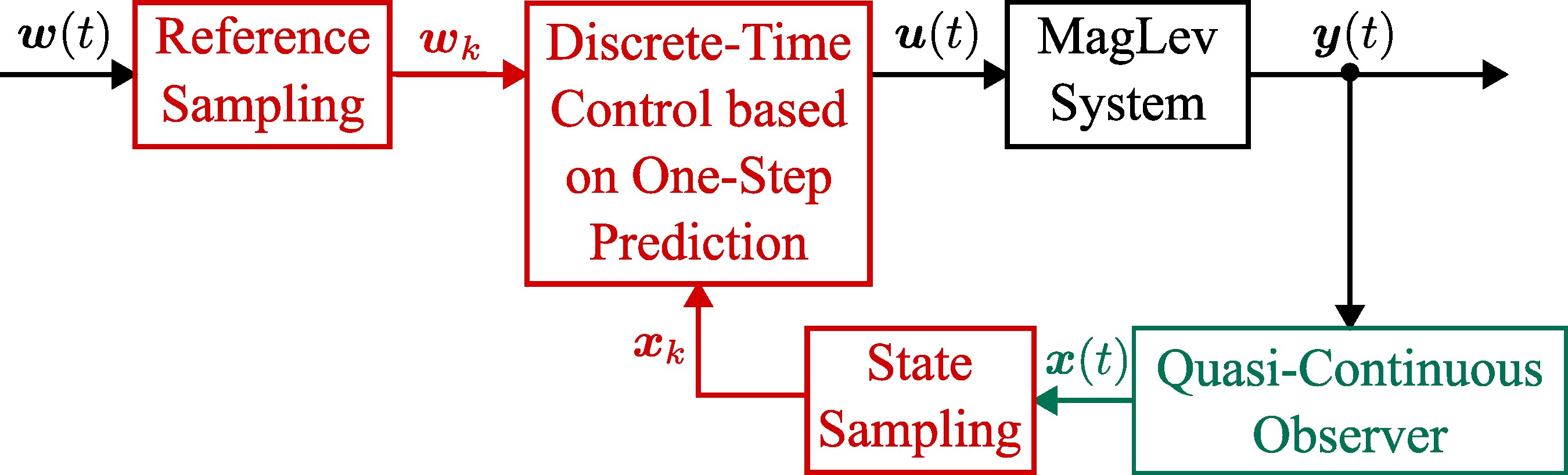}
		\caption{Schematic block diagram of the experiment. Continuous-time components are drawn in black, quasi-continuous components in green, and discrete-time components in red. $y(t)=\left[ s(t), i(t) \right]^{T} $}
		\label{fig:block_diagram}
	\end{center}
\end{figure}
We choose the control gains to be
\begin{equation*}
	C = m\lambda_{s} \lambda_{p}, \quad k_{1} = -m(\lambda_{s}+\lambda_{p}),\quad k_{2} = 80,
\end{equation*}
where $\lambda_{s}=\lambda_{p}=-50$ are the desired eigenvalues of the subsystem $[s,p]^{T}$. The momentum $p$ is not directly measurable, and is calculated using a quasi-continuous Luenberger observer operating at the base sampling time of $1\,\textrm{ms}$ as a proof of concept. Fig.~\ref{fig:block_diagram} shows the schematic block diagram for the conducted experiment. The desired position is switched between 2 setpoints $s^{\ast}$ and smoothed by a low-pass filter with $0.05\,\textrm{s}$ as the time constant.
\begin{figure}
	\begin{center}
		\includegraphics[width=8.4cm]{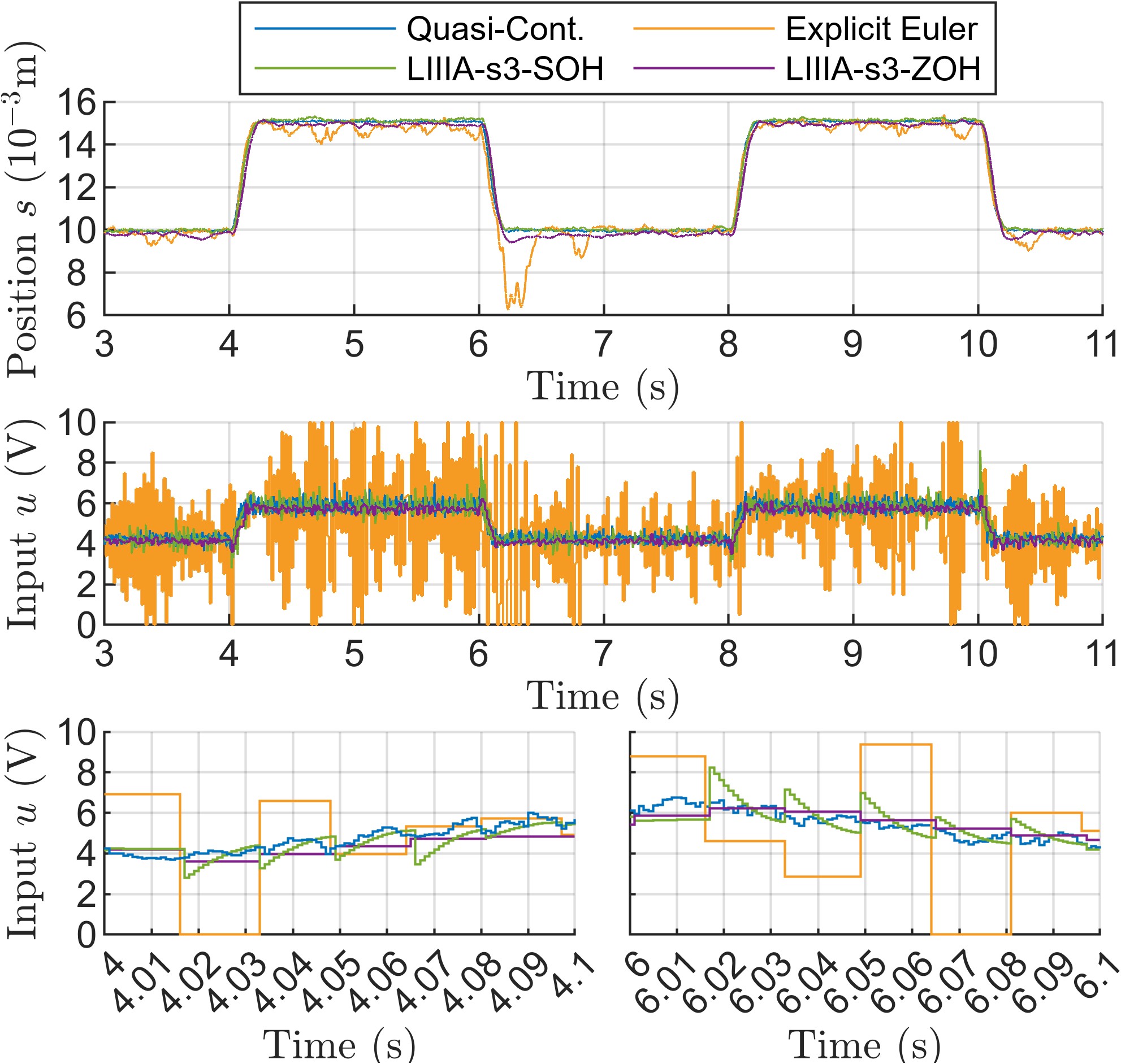}
		\caption{Experimental results for the implementation of 3-stage Lobatto IIIA collocation at $h=16\,\textrm{ms}$.}
		\label{fig:experiment_s3}
	\end{center}
\end{figure}

Through the experiments, the traditional emulation control (``Explicit Euler'') struggles to remain stable at a sampling time of $16\,\textrm{ms}$, while with the 3-stage Lobatto IIIA method, both controllers using second-order-hold input (``LIIIA-s3-SOH'') and piecewise constant input (``LIIIA-s3-ZOH'') can follow the setpoint with very high accuracy, as shown in Fig.~\ref{fig:experiment_s3}. The longest admissible sampling times of different implementations with our chosen control gains are listed in Table \ref{tb:sampling_limits}.

\begin{figure}
	\begin{center}
		\includegraphics[width=8.4cm]{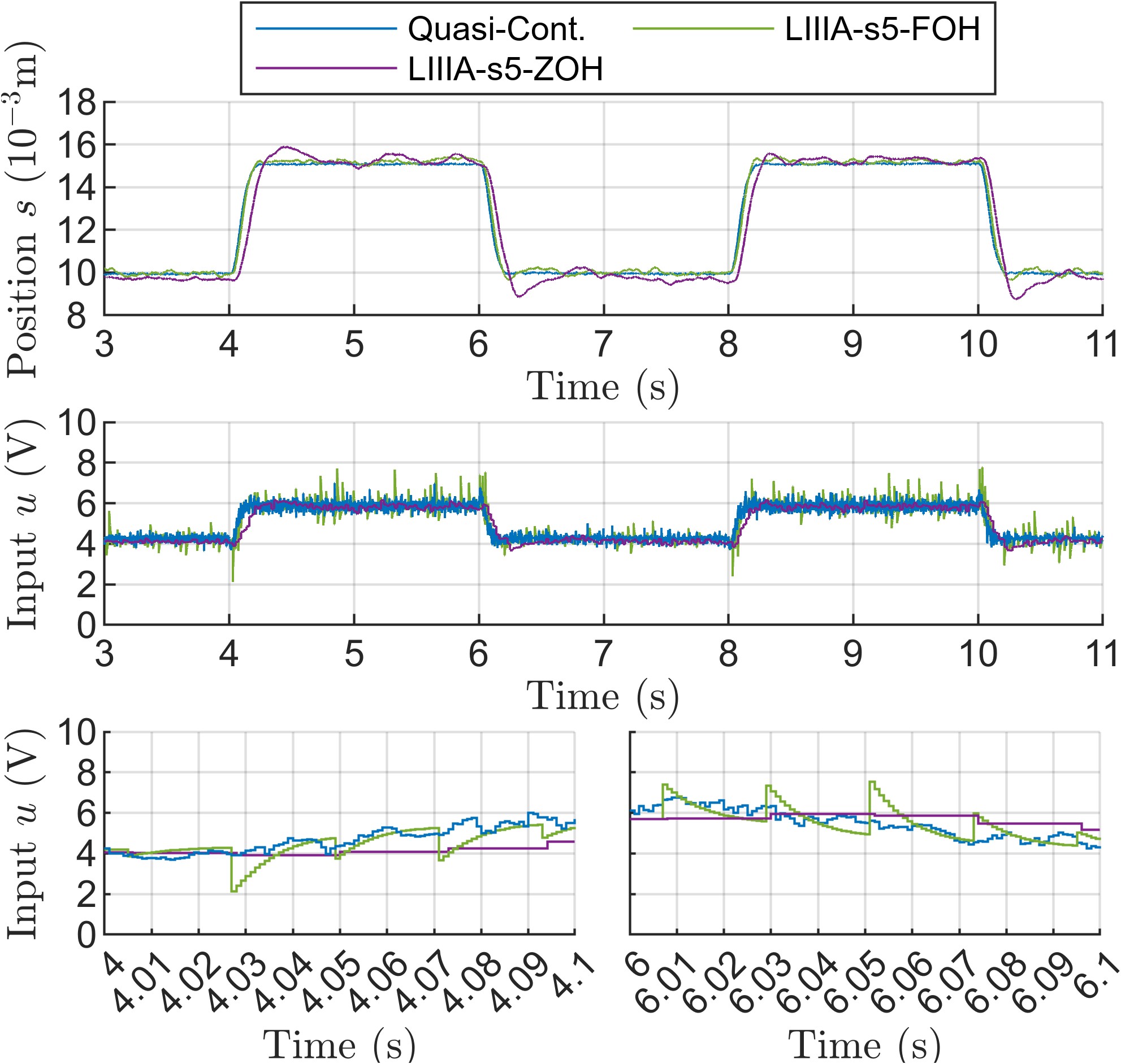}
		\caption{Experimental results for the implementation of 5-stage Lobatto IIIA collocation at $h=22\,\textrm{ms}$.}
		\label{fig:experiment_s5}
	\end{center}
\end{figure}

The times in Table~\ref{tb:sampling_limits} suggest that while higher order methods provide smaller approximation error, they are also more sensitive to noise and modeling errors. Therefore the 5-stage Lobatto IIIA method fails at a shorter sampling time than the 4-stage method. In addition, with the conversion to constant input, all the implementations regardless of orders, fail at similar sampling times (see Fig.~\ref{fig:experiment_s5} for 5-stage for example), indicating that the conversion reduces all the methods consistently to a certain lower order.

\begin{table}[bp]
	\begin{center}
		\caption{Limits of different implementations}\label{tb:sampling_limits}
		\begin{tabular}{ccc}
			Method &Shaped Input & Constant Input \\ \hline
			Explicit Euler & - &$16\,\textrm{ms}$ \\
			3-stage LIIIA &$38\,\textrm{ms}$ & $22\,\textrm{ms}$ \\
			4-stage LIIIA & $42\,\textrm{ms}$ & $23\,\textrm{ms}$ \\
			5-stage LIIIA & $34\,\textrm{ms}$ & $22\,\textrm{ms}$ \\ \hline
		\end{tabular}
	\end{center}
\end{table}

\section{Conclusions and Outlook}
\label{conclusions}
In this contribution, we generalized Lobatto IIIA collocation and Hermite interpolation of arbitrary order towards the Hermite-Obreschkoff methods. We proved that when the nodes are the zeros of the shifted Legendre polynomial, the combination of Hermite-Obreschkoff formulas with pairs $(n,m),\, n+m=s$ can be solved uniquely for arbitrary $s$, and the solution yields equivalently Lobatto IIIA collocation and Hermite interpolation. In this sense, these two methods are equivalent.

Through experimental results with a new IDA-PBC control law for the MagLev system, we showed that the implementation of higher-order Hermite-Obreschkoff methods can indeed reduce the model mismatch and improve the performance of the controller. The proposed conversion from higher-order-hold input to piecewise constant input limits the maximum achievable sampling times

We currently investigate the relation between our higher-order discrete-time control and single-horizon model predictive control, and explore the possibility of considering constraints, as well as the smoothness of the input signal. It is also our interest to find out if the $s$-stage $q$-derivative collocation methods generalized by \cite{kastlunger_wanner:1972} could also be derived using the HO formula.

\small
\bibliography{ifacconf}

\appendix

\end{document}